\documentclass[%
 reprint,
 amsmath,amssymb,
aip,
jcp
]{revtex4-1}

\usepackage{bm}
\usepackage{graphicx}

\newcommand*{\la}{\langle}
\newcommand*{\ra}{\rangle}

\newcommand{\avr}[1]{\langle\,  #1   \,\rangle}
\newcommand{\SAPT}[1]{$J_{\textrm{SAPT}} #1 $}

\usepackage{dcolumn}
\newcolumntype{d}[1]{D{.}{.}{#1}} 

\begin{document}

  \title{Exchange  splitting of the interaction energy 
  and~the~multipole expansion of~the~wave function }

\author{Piotr Gniewek}
 \email[]{pgniewek@tiger.chem.uw.edu.pl}
\author{Bogumi\l{} Jeziorski}
 \email[]{jeziorsk@chem.uw.edu.pl}

 \affiliation{Faculty of Chemistry, University of Warsaw, 
Pasteura 1, 02-093 Warsaw, Poland}

\date{\today}

\begin{abstract}
 The exchange splitting $J$  of the interaction  energy  of  the hydrogen atom with a proton  
 is calculated using the conventional surface-integral formula  $J_{\textrm{surf}}[\varphi]$,
 the volume-integral  formula  of  the symmetry-adapted perturbation theory
 $J_{\textrm{SAPT}}[\varphi]$, and a variational volume-integral formula  $J_{\textrm{var}}[\varphi]$. 
The calculations are based on the 
multipole expansion of the wave function $\varphi$, which is divergent for any internuclear distance $R$. 
Nevertheless, the resulting  approximations to the leading coefficient $j_0$ in the large-$R$ asymptotic series 
$J(R)  =  2 e^{-R-1} R ( j_0 + j_1  R^{-1} + j_2  R^{-2} +\cdots ) $ converge,   
with the rate   corresponding to  the convergence radii
equal to 4, 2, and 1 when the  $J_{\textrm{var}}[\varphi]$, $J_{\textrm{surf}}[\varphi]$, and 
 $J_{\textrm{SAPT}}[\varphi]$ formulas are used, respectively. 
Additionally, we observe that  also the higher $j_k$ coefficients are predicted correctly when the multipole expansion is used 
in the  $J_{\textrm{var}}[\varphi]$ and  $J_{\textrm{surf}}[\varphi]$ formulas. The SAPT formula  $J_{\textrm{SAPT}}[\varphi]$ 
predicts correctly only the first two coefficients, $j_0$ and $j_1$,  gives a wrong value of $j_2$, and diverges for higher $j_n$. 
Since the variational volume-integral formula can be easily generalized to many-electron systems 
and evaluated with standard basis-set techniques of quantum chemistry,  it provides an alternative for the determination 
of the exchange splitting   and the exchange contribution of the interaction potential in general. 
\end{abstract}


\maketitle

\section{Introduction}

Multipole expansion of the interaction energy is one of the cornerstones of the theory of atomic 
  interactions \cite{Stone:13, Jeziorski:94}. 
It is indispensable for the description 
of potential energy curves  in the domain of large interatomic 
separations $R$,   
where it approximates the interaction energy $E_{\textrm{int}}(R)$ 
asymptotically \cite{Ahlrichs:76,Morgan:80}  as a series  in the 
inverse integer powers of $R$,
\begin{equation}\label{eq1}
 E_{\textrm{int}}(R) \sim  \sum_n C_n R^{-n} ,
\end{equation}
 the coefficients   $C_n$ being usually referred to as the  van der Waals constants. 
The series on the r.h.s. of Eq.~(\ref{eq1})
cannot converge to $E_{\textrm{int}}(R)$, since the latter  
contains also exponentially decaying terms due to charge penetration 
and resonance tunneling (exchange) of electrons between interacting systems 
 \cite{Damburg:84,Cizek:86}.  
Moreover, the    multipole  expansion   (\ref{eq1}) is believed to be divergent, 
although  this has been rigorously proved only for H$_2^+$ i.e. for the interaction of a hydrogen atom with a proton
\cite{Damburg:84,Graffi:85,Cizek:86} and for the second order interaction energy 
of two hydrogen atoms \cite{Young:75}. 
One may note, however, that the multipole expansion  has been proved to   converge  for interactions 
of confined atoms \cite{Zhang:12} and in calculations with finite basis sets 
\cite{Jansen:00}, although the limits obtained in both these cases differ 
from the true interaction energy.

The multipole expansion (\ref{eq1}) is closely related to and can be obtained from the multipole expansion 
of the wave function \cite{Ahlrichs:76}  
\begin{equation}\label{eq2}
 \varphi\sim \varphi_0 + \sum_{n }  R^{-n} \varphi_n ,
\end{equation}
where $\varphi_0$ is the product of the wave functions 
of the non-interacting monomers, and $\varphi_n$'s are the 
multipole corrections to the wave function \cite{Ahlrichs:76}. 
The r.h.s. of Eq.~(\ref{eq2}) does not represent the asymptotic approximation 
of the exact wave function $\psi$ since it lacks the full  permutational  
and/or spacial symmetry of $\psi$. 
It has been shown, however,  that except for pathological cases \cite{Klein:87},
the correct asymptotic expansion  of the wave function can be  obtained by 
symmetry projection \cite{Ahlrichs:76}, i.e.,
\begin{equation}\label{eq3}
 \psi  = {\cal  A} \varphi_0 + \sum_{n=1}^N  R^{-n}{\cal  A}\varphi_n  
       + O(R^{-N-1}) ,
\end{equation}
where  ${\cal A}$ is the  projection operator imposing the appropriate   
symmetry of the state $\psi$.  Equation (\ref{eq3}) shows that Eq.~(\ref{eq2}) 
provides in fact the asymptotic expansion of  a genuine  primitive function 
$\varphi$, as defined, e.g., in Ref. \onlinecite{Kutzelnigg:80}.  
Similarly as  the expansion of Eq.~(\ref{eq1}), the multipole expansions 
for the wave function, Eq.~(\ref{eq3}), and for the primitive wave function, 
Eq.~(\ref{eq2}), are expected to be divergent  in the $   L ^2$ norm 
(the convergence of  the expansion for the wave function would imply 
the convergence for the interaction energy).
 
The interaction  energies of two or more states have the same asymptotic 
expansion when they  differ only by exponentially small exchange terms.   
For instance, for the H$_2^+$ ion,  for the  H$_2$ molecule and  alkali dimers,
 or for homonuclear ions with one electron outside  the closed shell, 
the interaction energies for the lowest \emph{gerade} and \emph{ungerade} 
states can be written as  
\begin{equation} \label{def_J} 
{\cal E}_{g,u}(R)  = Q(R)  \pm  J(R), 
\end{equation}
where $Q(R)$ is the so-called Coulomb energy, assumed to be well represented 
at large $R$ by the series (\ref{eq1}), and $J(R)$ is the so called exchange 
energy falling off exponentially with $R$  (the + and $-$ signs are used for  
the $g$ and  $u$ states, respectively).   The exchange energy  $J$ (or the 
exchange splitting $-2 J$)  is of paramount importance for the understanding 
of weak intermolecular interactions~\cite{Szalewicz:05},  chemical bonds 
or magnetism \cite{Herring:62}, and is relevant experimentally, as
it determines the rates of resonant charge exchange processes in slow
atomic   collisions  \cite{Galitski:81,Smirnov:01}. 

The exchange energy $J(R)$ and its large $R$ asymptotic expansion are much more 
difficult to calculate than the long-range part of the interaction energy, 
given by Eq.~(\ref{eq1}). This is due to the fact that $J(R)$, as a tunneling 
effect, is sensitive to the values of wave functions in the classically 
forbidden region of the configuration space, where the  wave function amplitudes are very 
small and are hard to determine accurately using  the conventional techniques 
of electronic structure theory. Instead of the wave functions $\psi_{g}$ and
$\psi_u$,  it is more convenient to work with a primitive function \cite{Kutzelnigg:80}  
$\varphi$, such that  $\psi_{g} ={\cal A}_{g}\varphi$,  
and   $\psi_{u} ={\cal A}_{u}\varphi$, where  ${\cal A}_{g}$, 
and ${\cal A}_{u}$ are appropriate symmetry projectors. When $\varphi$ is known, 
$J(R)$ can be obtained from the surface integral formula 
\cite{Firsov:51,Holstein:52,Herring:62},
\begin{equation}\label{eq:J_surf}
 J_{\textrm{surf}}[\varphi] = \frac{- \int_M \varphi \mathbf{\nabla} \varphi \textrm{d} 
\mathbf{S} }{ \la \varphi | \varphi \ra  - 2 \int_{\textrm{right}} \varphi^2 \textrm d V },
\end{equation}
where $M$ indicates the median plane of the molecule, and the volume integral 
in the denominator is taken over the half of the space right to $M$ ($\varphi$ 
is understood to be localized on the left of $M$). Atomic units are used in 
Eq.~(\ref{eq:J_surf}) and throughout the paper. The surface-integral method has 
been extended to hydrogen molecule \cite{Gorkov:64, Herring:64, Burrows:12} 
alkali-metal dimer cations 
\cite{Smirnov:65, Bardsley:75, Tang:92, Scott:02, Scott:04, Jamieson:09, Burrows:10}, 
neutral homo- and heterodimers \cite{Kleinekathoefer:96, Kleinekathoefer:00, Scott:04b, Yiu:11, Chen:14},
excited states of H$_2^+$ ion \cite{Scott:91}, and interactions of diatomic molecules with atomic ions \cite{Khoma:08}.
For the discussion of other extensions of this theory see Ref. \onlinecite{Chibisov:88}.

In our previous work \cite{Gniewek:14} 
we   presented a volume integral
formula for $J(R)$ rooted in the Symmetry Adapted Perturbation Theory
(SAPT),
\begin{equation}\label{SAPT}
 J_{\textrm{SAPT}}[\varphi] = \frac{ \la \varphi_0 | V P  \varphi \ra 
 \la \varphi_0 | \varphi \ra
- \la \varphi_0 | V \varphi \ra \la \varphi_0 | P  \varphi \ra }{
\la \varphi_0 | \varphi \ra^2
- \la \varphi_0 | P \varphi \ra^2 } ,
\end{equation}
where $V$ is the operator collecting Coulombic interactions of particles of one 
monomer with those of the other, and  $P $ is the operator inverting the electronic 
coordinates with respect to the midpoint of the internuclear axis (for H$_2^+$ and H$_2$) or permuting 
electrons between monomers (for larger systems). 
 When $\varphi$ is approximated via basis set expansions or via the expansion in powers 
of the interaction operator $V$,  the formula of Eq.~(\ref{SAPT}) was  shown to give  much better results 
 \cite{Gniewek:14}  than the surface-integral formula (\ref{eq:J_surf}). In fact, with $\varphi$ expanded in powers of $V$,   
the formula  (\ref{SAPT})  gives the expansion of the exchange energy in the symmetrized 
Rayleigh-Schr\"odinger (SRS) perturbation theory \cite{Jeziorski:78}, which
forms the basis for the calculation of exchange effects in most of the practical implementations of SAPT 
\cite{Szalewicz:12,Sherrill:12,Jansen:14,Hesselmann:14}. 

In this communication we consider another volume-integral 
formula, which is variational in its origin, and, as we shall show,  surpasses
$J_{\textrm{surf}}[\varphi]$ and $J_{\textrm{SAPT}}[\varphi]$ in accuracy,
\begin{equation}\label{eq:J_var}
 J_{\textrm{var}}[\varphi] = \frac{ \la \varphi | H P  \varphi \ra 
\la \varphi | \varphi \ra
- \la \varphi | H \varphi \ra \la \varphi | P  \varphi \ra }{
\la \varphi | \varphi \ra^2
- \la \varphi | P  \varphi \ra^2 } . 
\end{equation}
A similar expression and its simplified forms have been considered in the literature 
 \cite{Galitski:81,Janev:81,Chibisov:88,Khoma:08} in the theory of resonant and 
non-resonant atom-ion charge exchange processes (Landau-Zener theory)
but not in the present context of accurate \emph{ab initio} calculations of the exchange energy.  

The purpose of the present work is to investigate the performance of the 
surface-integral formula,  Eq.~(\ref{eq:J_surf}), and the two volume-integral formulas,
Eqs.~(\ref{SAPT}), and (\ref{eq:J_var}),  in the prediction of the exchange splitting 
energy $J(R)$ when the primitive function $\varphi$ is represented by the  multipole expansion,
of Eq.~(\ref{eq2}).  Since the error of the asymptotic series (\ref{eq3}) truncated after the $N$th  term 
is  the order of  $R^{-N-1}$, it is not obvious that this series  can be useful to correctly generate 
the exponentially small exchange terms in the energy. In fact is has been argued \cite{Harrell:80} 
that ``it is hopeless  to try to compute the splitting by any perturbative series expansion 
in $1/R$".  On the other hand, such a procedure, if successful, would be attractive computationally since 
the expansion (\ref{eq2}) is relatively easy to generate and Eq.~(\ref{eq3}) provides the simplest 
approximation to the wave function that is asymptotically correct in the whole configuration  space.  

The performance of different approximations of $J(R)$
can be best  investigated for the H$_2^+$ molecular 
ion.  For this simplest,  
archetypal system the exchange energy can be computed from the exactly known asymptotic expansion 
\cite{Ovchinnikov:65, Komarov:67, Damburg:68, Cizek:86}:
\begin{equation}\label{J_exp}
  J(R) = 2 e^{-R-1} R ( j_0 + j_1 /R + j_2 / R^2 + j_3 / R^3 + \cdots ) .
\end{equation}
\v{C}\'\i\v{z}ek et al. gave accurate values of the first 
52 $j_k$'s in Ref.~\onlinecite{Cizek:86}. For the interaction of two hydrogen atoms 
only the first term in the analogous expansion is known \cite{Gorkov:64,Herring:64}, although
even the functional form of this  leading  term has been debated recently \cite{Burrows:12}.  For larger 
systems only approximate form of the  leading  term is known 
\cite{Smirnov:65, Bardsley:75, Tang:92, Scott:02, Scott:04, Jamieson:09, Burrows:10,Kleinekathoefer:96,%
Kleinekathoefer:00, Scott:04b, Yiu:11, Chen:14}. Its accuracy is hard to ascertain since no reference data
sufficiently accurate at large  $R$ are available. Therefore, we performed our investigation  for the 
H$_2^+$ molecular ion and compared our results with the exact formula  of Eq.~(\ref{J_exp}). 

The problem considered  by us was first studied by Tang, Toennies, and Yiu \cite{Tang:91} who  evaluated 
the surface integral  (\ref{eq:J_surf}) with $\varphi$ represented by the multipole expansion of the 
Rayleigh-Schr\"odinger (RS) perturbation series in $V$ (the {\em polarization expansion} \cite{Jeziorski:94,Hirschfelder:67}).  
They were able to sum this series to infinite order and have shown 
that   the first, $j_0$ term in Eq.~(\ref{J_exp}) is obtained correctly in this way.  
They also obtained reasonably good approximate values of $j_1$ and $j_2$ by representing $\varphi$    
through the second order in $V$. These authors did not consider the alternative, volume-integral formulas. 

It may be noted that  for H$_2^+$  the exchange energy $J(R)$ can be approximately 
obtained directly from the series  (\ref{eq1}) 
without using the wave function expansion of Eq.~(\ref{eq2}).  
Brezin and Zinn-Justin \cite{Brezin:79}  have 
shown that the large $n$ behavior of the van der Waals constants $C_n$ in of Eq.~(\ref{eq1}) 
is related to the exchange energy via the relation
 \begin{equation}\label{BZ} 
 C_n  \approx  - \int_0^\infty \!\!\!     R^{n-1} [ J(R) ]^2\, \textrm d R  .
\end{equation}
Inserting the   expansion (\ref{J_exp}) into Eq.~(\ref{BZ}), performing integration  over $R$, and comparing  
with the known   $1/n$ expansion of $C_n$  \cite{Morgan:80,Cizek:80}, 
\begin{equation}\label{Cn}
 C_n    =  
  - \frac{1}{e^2} \frac{(n+1)!}{2^n} 
  \bigg ( 1 + \frac{2}{n} - \frac{20}{n^2} +   O\big(n^{-3}\big) \bigg ) .
\end{equation}
one finds the correct values of $j_0= -1 $ and $j _1= -1/2$, while for $j_2$ an incorrect value of  19/8 is obtained,  24\% \ smaller than the accurate value of $j_2$ equal to  
25/8.  Thus only the first two terms in the expansion (\ref{J_exp}) can be obtained exactly in this way.  This method 
requires the knowledge of an analytic form of  the $n$ dependence of $C_n$, which can hardly be expected to be available for larger systems.   

The organization of the paper is as follows. In Sec.~\ref{sec:theory} we define the multipole and polarization expansions 
 of the primitive function, discuss their applications to the evaluation of the asymptotics of the exchange energy and derive 
analytically the series representation for $j_0$ predicted by the SAPT's volume-integral formula. In Sec.~\ref{sec:numerical_results}  we present 
numerical results obtained with the large-order multipole expansion 
of the primitive function and with the first-order polarization wave function. A~successful application of a simple variational 
approximation to the primitive function is also presented. 
Finally, in Sec.~\ref{sec:conclusions}  we present conclusions of our investigation.

\section{\label{sec:theory} Theory}

\subsection{Primitive function and the variational 
volume-integral formula}

A primitive function $\varphi$ \cite{Kutzelnigg:80,Hirschfelder:67}
is a nontrivial linear combination of the asymptotically degenerate wave functions 
$\psi_g$ and $\psi_u$ of the \emph{gerade} and \emph{ungerade} states, 
\begin{equation}
 \varphi = c_1 \psi_g + c_2 \psi_u ,
\end{equation} 
 from which these exact states can be recovered by projection
\begin{equation}\label{psi_gu}
 \psi_g =      \tfrac{1}{2} (1+P) \varphi , \quad     \ \
 \psi_u =      \tfrac{1}{2} (1-P) \varphi .
\end{equation}
We shall also require that  $\varphi$ is a ``genuine  primitive 
 function" \cite{Kutzelnigg:80}, i.e., that  it  is  localized in the same way as $\varphi_0$.   
This means that $ \la \varphi  | P  \varphi \ra$ vanishes exponentially  or, more rigorously, that  
\begin{equation}\label{local}
 \la \varphi  | P   \varphi \ra = o(R^{-n})  
\end{equation}
for all  integers $n > 0$.

Substituting  the wave functions of the form given by  Eq.~(\ref{psi_gu}) in the  Rayleigh-Ritz
functional and taking into account that $[H,P]\!=\!0$ and   $(1\pm P)^2   \sim   1\pm  P$  one finds
\begin{equation}\label{RRdiff}
 E_g \!  -\!  E_u \! = \! \frac{ \la \varphi | H \varphi \ra\! +\! \la \varphi |H\!P \varphi \ra }
            { \la \varphi |\varphi \ra  \!+ \!\la \varphi |P \varphi \ra } 
\!-\!\frac{ \la \varphi | H  \varphi \ra\! - \!\la \varphi |H\! P  \varphi \ra }
            { \la \varphi|  \! \varphi \ra    \!-\la \varphi |P \varphi \ra } .            
\end{equation}
When the subtraction in Eq.~(\ref{RRdiff}) is carried out analytically, the $R$ independent and the long-range $R^{-n}$ terms in the numerator 
cancel out in view of Eq.~(\ref{local}) and one obtains Eq.~(\ref{eq:J_var})  for the exponentially vanishing energy splitting $J(R)$.  
The explicit form of the  non-relativistic Hamiltonian $H$ of  the hydrogen atom at  point $a$ interacting with a proton at point $b$ 
is $H=H_0+V$, where 
\begin{equation}\label{eq:H0_V_def}
 H_0 = -\frac{1}{2} \nabla^2 - \frac{1}{r_a}, 
  \quad\quad V = -\frac{1}{r_b} + \frac{1}{R} ,
\end{equation}
 $r_a$ and $r_b$ denoting  the electron-nucleus distances. The operator $V$ of 
 Eq.~(\ref{eq:H0_V_def})
appears   in the SAPT formula  of Eq.~(\ref{SAPT}).
 
\subsection{Multipole expansion of the primitive function}

The  multipole  expansion of the primitive function $\varphi$, Eq.~(\ref{eq2}), 
is obtained if the operator $V$  is represented as the 
sum of charge-multipole interactions  
\begin{align}\label{Vmult}
   V \sim  & \ R^{-2} \,V_2 + R^{-3}\, V_3  + R^{-4}\,  V_4+ \ldots , \\
           &   V_n  =  - r_a^{n-1} P_{n-1}(\cos \theta_a) ,
\end{align}
where $P_{n}(x)$ is the Legendre polynomial and $\theta_a$ is the polar angle 
at nucleus $a$.  The expansion of the eigenfunctions of $H_0+V$ in powers of $R^{-1}$ 
leads then to the following recurrence  equations  for the van der Waals constants 
$C_n$ and multipole corrections to the wave function $\varphi_n$:
\begin{equation} \label{A1}
  C_n = \sum_{m=2}^n \la \varphi_0 | 
 V_m \varphi_{n-m} \ra , 
\end{equation}
\vspace{-6ex}

\begin{equation}
 \label{A2} 
 (H_0 -  E_0) \varphi_n = 
 \sum_{m=2}^n  ( C_m - V_m ) \varphi_{n-m} ,
\end{equation}
where $E_0 = -\frac{1}{2}$  and $\varphi_0=\pi^{-1/2} e^{-r_a}$ are the 
ground-state energy and the wave function of the 
unperturbed hydrogen atom. To specify $\varphi_n$ uniquely we  assumed here the 
intermediate normalization  condition  of $\varphi$, i.e.,   $\la \varphi_0|\varphi \ra=1$ or, equivalently,  
  $\la \varphi_0|\varphi_n \ra =0$  for $n >0$. 

One can show by induction that $\varphi_n$ is a product of $\varphi_0$ and a  
polynomial  in $r_a$ and $\cos \theta_a$.  The functions $\varphi_2$ and $\varphi_3$ 
consist of only one partial wave each, $p$ and $d$, respectively, while 
the higher ones contain more than 
one partial-wave component.  
We calculated the van der Waals constants $C_n$ and wave function corrections 
$\varphi_n$  up to $n$=150 using a computer algebra program. 
We employed  the exact representations of rational numbers, 
so the calculated $C_n$ and $\varphi_n$ are  exact.

\subsection{\label{subsec:Polarization_expansion_phi} Polarization expansion of the primitive function }

Another useful approximation to the primitive function is obtained by a finite order 
polarization expansion\cite{Hirschfelder:67} (or polarization approximation). 
This is an application of the standard Rayleigh-Schr\"odinger perturbation theory 
to the Hamiltonian partitioning $H=H_0+V$, resulting in the expansion   
of $\varphi$ in powers of $V$,
\begin{equation}\label{polexp} 
\varphi =  \sum_{k=0}^{\infty} \varphi^{(k)} . 
\end{equation}
The individual terms in this expansion can be obtained from the    
equation 
\begin{equation}\label{poleqs}
 (H_0-E_0)\varphi^{(k)} = - V \varphi^{(k-1)} +  \sum_{j=1}^{k} E^{(j)} \varphi^{(k-j)} ,
\end{equation}
where $E^{(k)} = \la \varphi_0 | V \varphi^{(k-1)} \ra$,  and the recursive process  is initiated 
assuming  $\varphi^{(0)} = \varphi_0$ and $E^{(0)}=E_0$. To fully specify $\varphi^{(k)}$ 
we also assume  the intermediate normalization of $\varphi$, i.e.,   $\la \varphi_0|\varphi^{(k)} \ra =0$  for $k >0$. 
 For H$_2^+$ only the first-order term, $\varphi^{(1)}$, in the series  
(\ref{polexp}) is known in the closed form \cite{Dalgarno:57}. 

For systems like H$_2^+$ or H$_2$ the polarization expansion has been demonstrated numerically to converge  
to the wave function of the  ground, \emph{gerade}  state  \cite{Chalasinski:77,Tang:94}, although for large $R$ the 
convergence radius is only marginally  greater than unity \cite{Gniewek:14,Cwiok:92:pol,Kutzelnigg:92} and the convergence  
rate becomes prohibitively poor.  Thus, at infinite order the sum of the series 
(\ref{polexp}) does not represent a genuine primitive function since it does not satisfy 
neither the second Eq.~(\ref{psi_gu}) nor the locality condition  of Eq.~(\ref{local}).  
Nevertheless  the polarization expansion  (\ref{polexp})  provides an asymptotic approximation 
of the primitive function in the sense that \cite{Ahlrichs:76,Jeziorski:82}
\begin{equation}\label{eq:Phi_p_as_asympt_exp}
 \psi_\nu = {\cal A} \Phi^{(K)}+ O(R^{-\kappa(K+1)}) ,
\end{equation}
where $\Phi^{(K)}$  is the sum of the first $K+1$ terms  in Eq.~(\ref{polexp}),   
${\cal A}$ is a suitable symmetry projector,   $(1\pm P)/2$  in the case of H$_2^+$,  and $\kappa=2$, when at least one of interacting subsystems is charged  
and $\kappa=3$ otherwise. 

Each term in Eq.~(\ref{polexp})  can be represented by the multipole expansion in powers of $R^{-1}$,
\begin{equation}\label{phink}
 \varphi^{(k)}   \sim  \sum_{n }  R^{-n}\, \varphi^{(k)}_n .
\end{equation}  
The series on the r.h.s. of  (\ref{phink}) cannot converge in the whole configuration space 
but is expected to provide the large-$R$ asymptotic expansion of $\varphi^{(k)}$ in the 
$L^2$ norm.  
Dalgarno and Lewis \cite{Dalgarno:55}  have shown  that 
\begin{equation}\label{phi1}
  \varphi^{(1)}  \sim \sum_{n=2}^{\infty} R^{-n}     \left ( \frac{ r_a^{n-1} }{n-1} + \frac{ r_a^{n} }{n} \right ) \,  \varphi_0 \, P_{n-1}(\cos \theta_a) .
\end{equation}
For $k > 1$ the coefficients $ \varphi^{(k)}_n$, similarly as $ \varphi_n$,  are  products of polynomials in $r_a$ and $\cos \theta_a$ and the function $\varphi_0$.  
One can easily show that 
\begin{equation}\label{polexp_phi_n}
\varphi_n = \sum_{k=1}^{[n/2]}  \varphi^{(k)}_n,
 \end{equation}
where $[r]$ denotes the entire value  of $r$.  Eq.~(\ref{polexp_phi_n}) may be viewed 
as the expansion of $\varphi_n$ in powers of $V$, i.e., the polarization expansion
of $\varphi_n$.  We shall also need the multipole expansion of  $\Phi^{(K)}$ truncated 
after  the $R^{-N}$ term
\begin{equation}
\Phi^{(K)}_N  = \sum_{k=0}^{K} \sum_{n=0}^N \, R^{-n} \,\varphi^{(k)}_n .
 \end{equation}
 Note that  for $K\ge N/2$  we have $ \Phi_N^{(K)}=\Phi_N $, where  
\begin{equation}\label{PhiN}
 \Phi_N = \sum_{n=0}^N R^{-n} \varphi_n.
\end{equation}
is the  the multipole expansion of  Eq.~(\ref{eq2})
truncated after  the $R^{-N}$ term.

\subsection{\label{subsec:SAPT_expansion_of_j0}SAPT expansion of $j_0$} 

In this subsection we shall use  the SAPT formula  \SAPT[$\varphi$] and the multipole 
expansion for  $\varphi$, Eq.~(\ref{eq2}),  to analytically evaluate the leading $j_0$ term in
 Eq.~(\ref{J_exp}). 
It is easy to see that in evaluating $j_0$,  the second term in the numerator  of
 Eq.~(\ref{SAPT}) can be neglected  and the denominator can be replaced by 1. 
 Thus, in view of  Eq.~(\ref{eq2}), $j_0$ can be obtained by analyzing  the large $R$  
asymptotics  of  the expression $\avr{\varphi_0|VP\Phi_N} $  or its individual terms  
$R^{-n}\avr{\varphi_0|VP \varphi_n} $.  The function  $\varphi_n$ can be written 
as a product  of $\varphi_0$ and a polynomial $f_n$  in $r_a$ and $\cos \theta_a$,  
i.e.,  $\varphi_n = f_n \varphi_0$,  where $f_0=1$, $f_1=0$ and 
\begin{equation}\label{fn}
f_n(r_a,\cos\theta) =\sum_{l=0}^{n-1} \sum_{m=0}^{n} d_{nml}\, r_a^mP_l(\cos \theta_a) 
\end{equation}
for $n\ge 2$.      
Employing the prolate spheroidal coordinates  coordinates
 $\xi =  (r_a + r_b )/ R$ , 
$\eta = ( r_a - r_b )/ R $
 and integrating by parts one finds that   
 \begin{equation}\label{as_from} \begin{aligned}
 \la \varphi_0 | V & P \varphi_0 f_n \ra 
        =  \frac{1}{4 } R  e^{-R} \bigg [ \int_{-1}^1 v( 1 , \eta ) d \eta \\[1ex]
 & + R^{-1} \int_{-1}^1 \frac{ d v( \xi , \eta ) }{ d \xi } \bigg |_{\xi=1} d \eta + O(R^{-2}) \bigg ],
\end{aligned}\end{equation}
where
\begin{equation}\label{v_def}
 v(\xi, \eta) = \big ( \xi  + \eta  \big ) 
            (\xi - \eta -2) 
   \,  f_n \big ( \tfrac{R (\xi - \eta )}{2}  , \tfrac{ 1 - \xi \eta }{ \xi - \eta } \big ) .
\end{equation}
Eqs.~(\ref{as_from}) and (\ref{v_def})  show that the leading  asymptotic term  of  \SAPT[$ f_n\varphi_0$]   depends only on the values of $f_n$ 
for $\xi=1$, i.e. on the values of $f_n$ on the line joining the nuclei $a$ and $b$. 
On this line $   \theta_b=0$ and   $ P_l(\cos\theta_b)=1$. 
Thus, the sought-after 
asymptotics in unchanged if the function $f_n(r_a,\cos\theta)$ is replaced by
\begin{equation} \label{fn1}
f_n(r_a,1)   =  \sum_{m=0}^{n} d_{nm }\, r_a^m ,
\end{equation}   
where 
\begin{equation}\label{dnm} 
d_{nm} =\sum_{l=0}^{n-1}   d_{nml}.
\end{equation} 
By inspecting Eqs.~(\ref{as_from}) and (\ref{v_def})  one can also see that 
the leading contribution  to $ \la \varphi_0   | V P \varphi_0 f_n \ra $ comes 
from the highest, $m=n$, power in Eq.~(\ref{fn1}). Thus, to evaluate the leading 
asymptotics of    \SAPT[$ f_n\varphi_0$]  one  can replace the  complete expression 
for $f_n$ by $d_n r_a^n$, where $d_{nn}$ is denoted by $d_n$   for brevity.

Inserting Eq.~(\ref{fn}) into Eq.~(\ref{A2}), setting $\theta =0$ and comparing coefficients   
at the highest power of $r_a$ one obtains the following recurrence relation
\begin{equation}\label{dn_rec}
n d_n = d_0+d_1 +d_2 + \cdots + d_{n-2} .
\end{equation}
Using Eq.~(\ref{dn_rec}) one can easily prove that
\begin{equation}
 d_{n}-d_{n-1} = - \frac{1}{n} \big ( d_{n-1} - d_{n-2} \big ) . 
\end{equation}
Taking also into that $d_0=1$ and $d_1=0$  one obtains
\begin{equation}
 d_{n}-d_{n-1} = \frac{ (-1)^n }{ n! } ,
\end{equation}
and, consequently,
\begin{equation} \label{dn} 
d_n = \sum_{m=0}^n \frac{(-1)^m}{m!} .
\end{equation} 
  In view of Eqs.~(\ref{as_from}) and (\ref{v_def}) the leading asymptotics of 
$ \la \varphi_0   | V P   r_a^n\varphi_0 \ra $ is given by 
\begin{equation}\label{as1}
\frac{-   R^{n+1} e^{-R}}{2^{n +2}}\!\! \int_{-1}^1(1+\eta)^2(1-\eta)^{n} d\eta 
 = \frac{-4  R^{n+1} e^{-R} }{(n+1)(n+2)(n+3)}. 
 \end{equation}
 Summing up  contributions from $\varphi_0$ up to $\varphi_N$ we find that the large $R$ asymptotics  
 of \SAPT[$\Phi_N$] is given by 
 \begin{equation}\label{j0_SAPT}
 \sum_{n=0}^N\, \frac{-4  d_n  }{(n+1)(n+2)(n+3)}\,R e^{-R}. 
  \end{equation}
Since the $d_n$ coefficients converge quickly to $1/e$  it is clear that the series  
 (\ref{j0_SAPT}) converges, although slowly, with the d'Alembert ratio
 $|a_{n+1}/a_{n}|$   equal $1- 3 n^{-1} + O(n^{-2})$,
 i.e., with the corresponding convergence radius equal to unity. 
To evaluate  its limit one can use the obvious identity 
\begin{equation}\label{ident}
p_{n}(q_{n+1} -q_{n}) =   p_{n+1}q_{n+1} -p_{n}q_{n} -  (p_{n+1}-p_{n})q_{n+1}, 
\end{equation} 
\noindent 
valid for arbitrary sequences $p_n$ and $q_n$. Setting 
 \begin{equation} \label{set}
 p_n=d_n,  \quad\quad  q_n= \frac{2}{(n+1)(n+2)},
 \end{equation}  
and summing both sides of Eq. ({\ref{ident}) over $n$ from $n=0$ up to $n=N$
we find 
\begin{equation}\begin{aligned} 
  \sum_{n=0}^N& \frac{-4  d_n }{(n+1)(n+2)(n+3)}  =\\
&   =   \frac{2d_{N+1}}{(N+2)(N+3)} - 1 -2\sum_{n=0}^N \frac{(-1)^{n-1}}{(n+3)!} , 
\end{aligned}\end{equation}  
where we took into account that $d_0=1$, $q_0=1$ and $d_{n+1} -d_{n} =(-1)^{n+1}/(n+1)!$. 
When $N\rightarrow \infty$ the first term on the r.h.s. vanishes and the last 
one has the limit equal to    $-2/e +1$. Thus, the series on  the l.h.s. converges 
to $2/e$  and one finds that \SAPT[$\Phi_N$] gives the correct asymptotics of $J(R)$ 
corresponding to $j_0=-1$. Having in mind that  the multipole expansion of the wave 
function,  Eq. (\ref{eq2}),  leads to the  divergent expansion for the total interaction energy
 we find it quite  remarkable  that,  when inserted into the SAPT formula \SAPT[$\phi$], 
this expansion  gives the  convergent series for such a subtle effect as the asymptotic 
exchange splitting  $(-2/e)Re^{-R}$.

\section{\label{sec:numerical_results} Numerical Results}

\subsection{Exchange splitting from the  multipole expansion 
of the primitive function}

Since the functions $f_n$ are polynomials in $r_a$ and $\cos\theta_a$, the exchange 
splittings $J_{\textrm{SAPT}}[\Phi_N]$, $J_{\textrm{surf}}[ \Phi_N]$ 
and $J_{\textrm{var}} [ \Phi_N ]$ can be obtained in a closed form for a wide range 
of $N$ employing computer algebra software. For instance, for $N=3$ we obtained
\begin{equation}\label{Jsurf3}
 J_{\textrm{surf}}[ \Phi_3 ]  =R e^{-R}    (
    - \tfrac{49 }{72} - \tfrac{7}{16}  \tfrac{1}{R}+ \tfrac{613}{384} \tfrac{1}{R^2}  
    + \cdots)  ,
\end{equation}
\begin{equation}\label{JSAPT3}
 J_{\textrm{SAPT}}[ \Phi_3 ]  =R e^{-R} \big  (
    -  \tfrac{32  }{45} - \tfrac{7}{30}\tfrac{1}{R} + \tfrac{97}{60  }  \tfrac{1}{R^2}
    +\cdots ),
\end{equation}
\begin{equation}\label{Jvar3}
 J_{\textrm{var}}[ \Phi_3 ]=R e^{-R} \big  (
    - \tfrac{4147 }{5670} - \tfrac{1369}{3780} \tfrac{1}{R} + \tfrac{17239}{7560}  \tfrac{1}{R^2}  +\cdots ).
\end{equation}
Comparing with Eq. (\ref{J_exp}) we see that after 
multiplication by $e/2$ the successive coefficients  at $1/R^k$  on  the r.h.s. 
of Eqs.~(\ref{Jsurf3})-(\ref{Jvar3})  represent   approximations to  the  $j_k$  
coefficients in the expansion (\ref{J_exp}).   These coefficients computed using the function 
$\Phi_N$  and the appropriate energy expressions will be denoted by
$j_k^{\textrm{surf}}[\Phi_N]$,   $j_k^{\textrm{SAPT}}[\Phi_N]$, and 
$j_k^{\textrm{var}}[\Phi_N]$.  In view of Eqs.~(\ref{Jsurf3})-(\ref{Jvar3}),  
for $N=3$  the corresponding approximations to the $j_0$ coefficient   are $-49e/144$,   $-16e/45$, and 
$-4147e/11340$,  and differ from the exact value   $j_0=-1$ by  7.5\%,  3.3\%, and 0.6\%, respectively.

In  Tables~\ref{tab:j0_vs_N}, \ref{tab:j1_vs_N},  and \ref{tab:j2_vs_N}
we show how the values of $j_0$, $j_1$, and $j_2$  computed from $\Phi_N$ 
converge when $N$  increases  (the rows for $N=1$   are absent since $\varphi_1=0$ so that  $\Phi_1=\Phi_0$). 
 For the constants $j_0$, $j_1$, and $j_2$  the fastest convergence by far is 
 observed in the case of the variational formula, whereas
 the SAPT formula gives the slowest convergence. 
 Moreover, the sequence $j_2^{\textrm{SAPT}}[\Phi_N]$ converges to a spurious value 
 of $55/24 \approx 2.292$, instead of the correct one, equal to 25/8 = 3.125.  We also 
 observed that for   $k \ge 3$ the sequences  $j_k^{\textrm{SAPT}}[\Phi_N]$ 
  diverge  while the sequences generated by the surface-integral  and variational formulas appear to be convergent for all constants $j_k$ that we computed.  In the case of the variational formula this convergence is demonstrated  numerically for $k \le 6$ 
 in Table~\ref{tab:jkvar}. It can be seen that the variational volume formula provides excellent accuracy also for further terms in the expansion of Eq.~(\ref{J_exp}).

\begin{table}[tb]
\caption{\label{tab:j0_vs_N} Values of $j_0$ calculated from the 
surface- and volume-integral formulas with  the  multipole 
expansion of  Eq.~(\ref{eq2}) truncated after the $R^{-N} \varphi_N$ term.  }
\begin{ruledtabular}
\begin{tabular}{ c d{2.8} d{2.8} d{2.11}  } 
$N$  & 
\multicolumn{1}{c}{$-j_0^{\textrm{SAPT}}$}   & 
\multicolumn{1}{c}{$-j_0^{\textrm{surf}}$}   & 
\multicolumn{1}{c}{$-j_0^{\textrm{var}}$}    \\
\hline
   0 &  0.9061    &   0.6796    &   0.9060\ 9394   \\ 
   2 &  0.9514    &   0.8601    &   0.9805\ 2309   \\ 
   3 &  0.9665    &   0.9250    &   0.9940\ 6656   \\ 
   4 &  0.9762    &   0.9625    &   0.9984\ 1193   \\ 
   5 &  0.9821    &   0.9811    &   0.9995\ 7106   \\ 
   6 &  0.9861    &   0.9905    &   0.9998\ 8567   \\ 
   7 &  0.9889    &   0.9953    &   0.9999\ 6973   \\ 
   8 &  0.9909    &   0.9976    &   0.9999\ 9204   \\ 
   9 &  0.9924    &   0.9988    &   0.9999\ 9791   \\ 
  10 &  0.9936    &   0.9994    &   0.9999\ 9946   \\ 
$\infty$ & 1.0  &   1.0    &   1.0  \\
\end{tabular}
\end{ruledtabular}
\end{table}

\begin{table}[tb]
\caption{\label{tab:j1_vs_N} Values of $j_1$ calculated from the 
surface- and volume-integral formulas with the multipole 
expansion of  Eq.~(\ref{eq2}) truncated after the $R^{-N} \varphi_N$ term.  }
\begin{ruledtabular}
\begin{tabular}{ c d{3.5} d{3.6} d{3.12}  } 
$N$  & 
\multicolumn{1}{c}{$-j_1^{\textrm{SAPT}}$}   & 
\multicolumn{1}{c}{$-j_1^{\textrm{surf}}$}   & 
\multicolumn{1}{c}{$-j_1^{\textrm{var}}$}    \\
\hline
   0 &  0.0      &   0.0     &   0.0          \\ 
   2 &  0.272   &   0.573  &   0.465\ 99   \\ 
   3 &  0.317   &   0.595  &   0.492\ 24   \\ 
   4 &  0.362   &   0.632  &   0.503\ 58   \\ 
   5 &  0.388   &   0.612  &   0.502\ 63   \\ 
   6 &  0.407   &   0.585  &   0.501\ 28   \\ 
   7 &  0.420   &   0.559  &   0.500\ 52   \\ 
   8 &  0.430   &   0.539  &   0.500\ 19   \\ 
   9 &  0.438   &   0.525  &   0.500\ 07   \\ 
  10 &  0.445   &   0.515  &   0.500\ 02   \\ 
$\infty$ & 0.5  &   0.5    &   0.5  \\
\end{tabular}
\end{ruledtabular}
\end{table}

\begin{table}[tb]
\caption{\label{tab:j2_vs_N} Values of $j_2$ calculated from the 
surface- and volume-integral formulas with the multipole 
expansion of  Eq.~(\ref{eq2}) truncated after the $R^{-N} \varphi_N$ term. 
The exact value of  $j_2$ is 25/8.   The sequence 
$j_2^{\textrm{SAPT}}[\Phi_N]$ converges to the incorrect value  equal to   $55/24\approx 2.292$. }
\begin{ruledtabular}
\begin{tabular}{ c d{3.5} d{3.6} d{3.12}  } 
$N$  & 
\multicolumn{1}{c}{$j_2^{\textrm{SAPT}}$}   & 
\multicolumn{1}{c}{$j_2^{\textrm{surf}}$}   & 
\multicolumn{1}{c}{$j_2^{\textrm{var}}$}    \\
\hline
      0 &  1.359  &   0.0     &   1.359\ 14 \\ 
       2 &  1.925  &   1.083  &   2.626\ 05 \\ 
       3 &  2.197  &   2.170  &   3.099\ 24 \\ 
       4 &  2.218  &   2.666  &   3.147\ 05 \\ 
       5 &  2.254  &   3.099  &   3.156\ 90 \\ 
       6 &  2.269  &   3.318  &   3.146\ 38 \\ 
       7 &  2.278  &   3.409  &   3.136\ 55 \\ 
       8 &  2.284  &   3.417  &   3.130\ 49 \\ 
       9 &  2.288  &   3.384  &   3.127\ 39 \\ 
      10 &  2.291  &   3.334  &   3.125\ 97 \\ 
$\infty$ &  2.292  &   3.125  &   3.125  \\
\end{tabular}
\end{ruledtabular}
\end{table}

The convergence of the sequence $j_0^{\textrm{SAPT}}[\Phi_N]$, shown in Table~\ref{tab:j0_vs_N},
to the correct limit is proved in Sec.~\ref{subsec:SAPT_expansion_of_j0}, while for the sequence  
 $j_0^{\textrm{surf}}[\Phi_N]$ the proof of convergence can be found in Ref. \onlinecite{Tang:91}. 
A rigorous mathematical  proof that the sequence   $j_0^{\textrm{var}}[\Phi_N]$, shown in the last column
of Table~\ref{tab:j0_vs_N}, converges to the correct limit is more  complicated     \cite{Gniewek:15}
and is beyond the scope of the present communication.
The convergence of the sequences presented in Tables \ref{tab:j1_vs_N} and \ref{tab:j2_vs_N}, as well as their  
limits have  been established numerically based on the calculations for very high values of $N$. 

\vspace{0.1cm} 

\begin{widetext}

\begin{table}[htb]
\caption{ \label{tab:jkvar}
Values of the coefficients $j_k$, $k\le 6$,  calculated from the variational formula and the 
  multiple expansion of the primitive wave function, Eq. (\ref{eq2}), truncated after $n=10$, 20, and 30.  
Data are rounded to show two digits differing from the exact value. 
The exact values are taken from Ref. \onlinecite{Cizek:86}.}
\begin{ruledtabular}
\begin{tabular}{ c d{3.9} d{3.15} d{3.21} d{4.13}  } 
 $k$  & 
\multicolumn{1}{c}{$j_k^{\textrm{var}}[\Phi_{10}]$} & 
\multicolumn{1}{c}{$j_k^{\textrm{var}}[\Phi_{20}]$} & 
\multicolumn{1}{c}{$j_k^{\textrm{var}}[\Phi_{30}]$} & 
\multicolumn{1}{c}{$j_k$} \\ 
\hline
0 & 
-0.999\ 999\ 46 &
-0.999\ 999\ 999\ 999\ 30 &
-0.999\ 999\ 999\ 999\ 999\ 999\ 20 & 
-1. \\ 
1 & 
-0.500\ 022 &
-0.500\ 000\ 000\ 13 &
-0.500\ 000\ 000\ 000\ 000\ 35 &
-0.5 \\
2 & 
3.125\ 97 &
3.125\ 000\ 022 &
3.125\ 000\ 000\ 000\ 13 &
3.125  \\
3 & 
2.708 &
2.729\ 164\ 0 &
2.729\ 166\ 666\ 628 &
2.729\ 166\ 666\ 667 \\
4 & 
10.46 &
10.216\ 39 &
10.216\ 145\ 842 &
10.216\ 145\ 833 \\
5 & 
38.9 &
37.847 &
37.864\ 321\ 3 &
37.864\ 322\ 9\\
6 & 
65. & 
114.1 &
113.263\ 89 &
113.263\ 65 \\
\end{tabular}
\end{ruledtabular}
\end{table}

\end{widetext}

To characterize the convergence rates of the sequences shown in Tables~\ref{tab:j0_vs_N}-\ref{tab:j2_vs_N}, we computed the  
increment  ratios $\rho_N(j^{\textrm{SAPT}}_k)$, $\rho_N(j^{\textrm{surf}}_k)$, and $\rho_N(j^{\textrm{var}}_k)$   defined as 
\begin{equation}\label{rhoN}
 \rho_N(j^{\textrm{SAPT}}_k) = - \frac{ j_k^{\textrm{SAPT}}[\Phi_{N-1}]-j_k^{\textrm{SAPT}}[\Phi_{N }] }{ j_k^{\textrm{SAPT}}[\Phi_{N+1}]-j_k^{\textrm{SAPT}}[\Phi_N]  }
\end{equation}
 and similarly for $\rho_N(j^{\textrm{surf}}_k)$,    and $\rho_N(j^{\textrm{var}}_k)$.

Since $ j_k^{\textrm{SAPT}}[\Phi_{N }] $ is a linear functional of $\Phi_{N }$, the increment ratio 
$  \rho_N(j^{\textrm{SAPT}}_k)$ is equal to the  inverse of the d'Alembert ratio 
 $ j_k^{\textrm{SAPT}}[\varphi_{N+1}]\,/\,j_k^{\textrm{SAPT}}[\varphi_N] $ and 
its limit when $N\rightarrow \infty$ determines the convergence radius of the series 
$\sum_n j_k^{\textrm{SAPT}}[\varphi_n]$. The surface-integral and variational formulas 
are nonlinear functionals of $\Phi_N$ but also in these cases the increment rations 
$\rho_N(j^{\textrm{surf}}_k)$ and $  \rho_N(j^{\textrm{var}}_k)$ can be interpreted as   
inverses of the d'Alembert ratios for the series with the coefficients  
$a_N^{\textrm{surf}} =  j_k^{\textrm{surf}}[\Phi_{N+1 }] - j_k^{\textrm{surf}}[\Phi_{N }]$ 
and similarly for $a_N^{\textrm{var}}$. We shall use the convergence radii of these series, given by the  
$N\rightarrow \infty$ limit of the corresponding increment ratios,
to characterize the convergence of the sequences  $j_k^{\textrm{surf}}[\Phi_{N }]$  
and $j_k^{\textrm{var}}[\Phi_{N }]$.

By least-squares fitting to the data of our numerical calculations we obtained the following large $N$ 
representation  of $\rho_N(j_k^{\textrm{var}}) $, $k=0,1,2,3$:
\begin{equation}\label{eq:rN_jk_var}\begin{aligned}
 \rho_N(j_0^{\textrm{var}}) & = 4 -\, 2 \,\, N^{-1} + O(N^{-2})  , \\
\rho_N(j_1^{\textrm{var}}) & = 4 - 10 \,N^{-1} +O(N^{-2})  , \\
\rho_N(j_2^{\textrm{var}}) & = 4 - 18 \, N^{-1} + O(N^{-2})  , \\
\rho_N(j_3^{\textrm{var}}) & = 4 -  26 \,N^{-1} + O(N^{-2})  , \\
\end{aligned}\end{equation}
where the relative uncertainties of the coefficients at the leading and subleading terms 
are of the order of 10$^{-20}$ or smaller, based on analysis of $j_k^{\textrm{var}}[\phi_N]$ for $N$ up to 60.

In the case of the surface integral formula we obtained 
\begin{equation}\label{eq:rN_surf}\begin{aligned}
 \rho_N(j_0^{\textrm{surf}}) & = 2 - \frac{3}{4\sqrt{e}}\, 2^{-N} + O(4^{-N}) , \\
\rho_N(j_1^{\textrm{surf}}) & = 2 -  4 \,N^{-1} +O(N^{-2})  , \\
\rho_N(j_2^{\textrm{surf}}) & = 2 -  8 \,N^{-1} +O(N^{-2})  , \\
\rho_N(j_3^{\textrm{surf}}) & = 2 -  12 \,N^{-1} +O(N^{-2})  . \\
\end{aligned}\end{equation}
The uncertainties of the coefficients in the above formulas are at most of the order of  10$^{-20}$, based on
analysis of $j_k^{\textrm{surf}}[\phi_N]$ for $N$ up to 150.
It is noteworthy   that the analytic form of  the   large $N$ representation  of $ \rho_N(j_k^{\textrm{surf}})$ 
is different for  $k=0  $ and for the higher  values of $k$.  We were able to verify the formulas 
for $\rho_N(j_0^{\textrm{surf }})$ and $\rho_N(j_0^{\textrm{var}})$ (obtained initially by fitting) 
by a rigorous mathematical derivation \cite{Gniewek:15}. 

Eqs. (\ref{eq:rN_jk_var}) and (\ref{eq:rN_surf}) show that the series of approximations to the $j_k$ coefficients 
obtained by using the multipole expansion in the surface-integral and variational volume-integral  formulas converge with  
the convergence radii equal to 2 and 4, respectively.  This is consistent with  the results presented in Tables~\ref{tab:j0_vs_N}-\ref{tab:j2_vs_N}, 
much faster convergence of the variational expression.  
Although the convergence radius of the series of approximations of $j_k$ is  independent of $k$, Eqs. (\ref{eq:rN_jk_var}) and (\ref{eq:rN_surf})   
show that  the rate of convergence deteriorates somewhat with the increase of $k$, in agreement with
the data presented in Tables~\ref{tab:j0_vs_N}-\ref{tab:jkvar}. 

When the volume-integral formula of SAPT is used one obtains the following inverse d'Alembert ratios for the series approximations to 
$j_k$:
\begin{equation}\label{eq:rN_jk_SAPT}\begin{aligned}
 \rho_N(j_0^{\textrm{SAPT}}) & = 1+\, 3 \,\, N^{-1} + O(N^{-2})  , \\
\rho_N(j_1^{\textrm{SAPT}}) & = 1+ 2 \,N^{-1} +O(N^{-2})  , \\
\rho_N(j_2^{\textrm{SAPT}}) & = 1+ 2 \, N^{-1} + O(N^{-2})  , \\
\rho_N(j_3^{\textrm{SAPT}}) & = 1+  \tfrac{26}{5} \,N^{-3} + O(N^{-4})  . \\
\end{aligned}\end{equation}
The first of these formulas is a direct consequence of the explicit expression for 
$j_0^{\textrm{SAPT}}[\Phi_N]$, given in Eq. (\ref{j0_SAPT}), while the remaining ones were obtained by 
least-squares fitting with uncertainties amounting to at most 10$^{-30}$, based on analysis of 
$j_k^{\textrm{SAPT}}[\phi_N]$ for $N$ up to 150.
 
Eq.~(\ref{eq:rN_jk_SAPT}) shows that the convergence radius of the SAPT expansion of the coefficients $j_0$,\ldots,$j_3$,
is equal to 1. In such a case the d'Alembert ratio test is inconclusive, but we can 
use  the Gauss criterion \cite{Arfken:68} to find out whether the  series converges or not.  
 According to this criterion  a series $a_0 + a_1 + a_2 + \ldots$ 
with $a_n > 0$  and the  coefficients behaving at large $n$ such that   
\begin{equation}
   \frac{ a_n }{ a_{n+1} }  = 1 +   h \,  n^{-1}  + O(n^{-p })  
\end{equation}
with $p > 1$, converges if $h > 1$ and diverges otherwise. 
We therefore can conclude that the $N \rightarrow \infty$ limits of  the sequences  $j_0^{\textrm{SAPT}}[\Phi_N]$, 
$j_1^{\textrm{SAPT}}[\Phi_N]$, and $j_2^{\textrm{SAPT}}[\Phi_N]$ do exist, 
while the sequence $j_3^{\textrm{SAPT}}[\Phi_N]$  diverges.  We  also  found that for $k\ge 4$ 
the sequences   $j_k^{\textrm{SAPT}}[\Phi_N]$ are   divergent since  in this case 
$\rho_N(j_k^{\textrm{SAPT}})$ is smaller than 1  at large $N$   (or $h <1$ if the Gauss test is applied).

In  Figure~\ref{fig:log10_Del_j0_vs_N}   we compare  the convergence of the    $j_0$  term of the asymptotic expansion of $J(R)$ obtained 
when the multipole expansion of the  primitive function is used in the variational, surface-integral, or SAPT formulas.
The regular behavior of relative errors allows   for  the use of extrapolation 
techniques. We used the  Levin  $u$-transform \cite{Levin:72} 
to extrapolate the results to the $N \rightarrow \infty$ limit. 
This method accelerates the convergence of partial sums 
$z_n = a_0 + a_1 + \ldots + a_n$ with a transformed sequence 
\begin{equation}
 u_n = \frac{ \sum_{i=0}^n (-1)^i \binom{ n }{ i } (i+1)^{n-2}
  z_i  a_i^{-1} }{ \sum_{i=0}^n (-1)^i \binom{ n }{ i } (i+1)^{n-2}
  a_i^{-1} } ,
\end{equation}
which under certain assumptions has the same limit as $z_n$ \cite{Weniger:89}.
Relative errors of $j_0$ obtained via eight-term Levin $u$-transform are 
presented in Fig.~\ref{fig:log10_Del_j0_vs_N} using lines with filled squares.
In case of $j_0^{\textrm{var}}$ and $j_0^{\textrm{surf}}$ eight-term
Levin $u$-transform increases the accuracy by about 4    and 3.5    orders of magnitude, 
respectively, for $N > 15$. On the other hand this method based on just eight terms 
seems unable to accelerate the convergence of $j_0^{\textrm{SAPT}}$
for $N > 12$. We found, however, that when more terms are used,  a significant decrease 
of error is possible. For instance, extrapolation from  150 values 
of $j_0^{\textrm{SAPT}}$ gives a result which differs from 
the exact value of $j_0 = -1$ by only   $\sim\! 10^{-38}$.

\begin{figure}[t!]
  \includegraphics{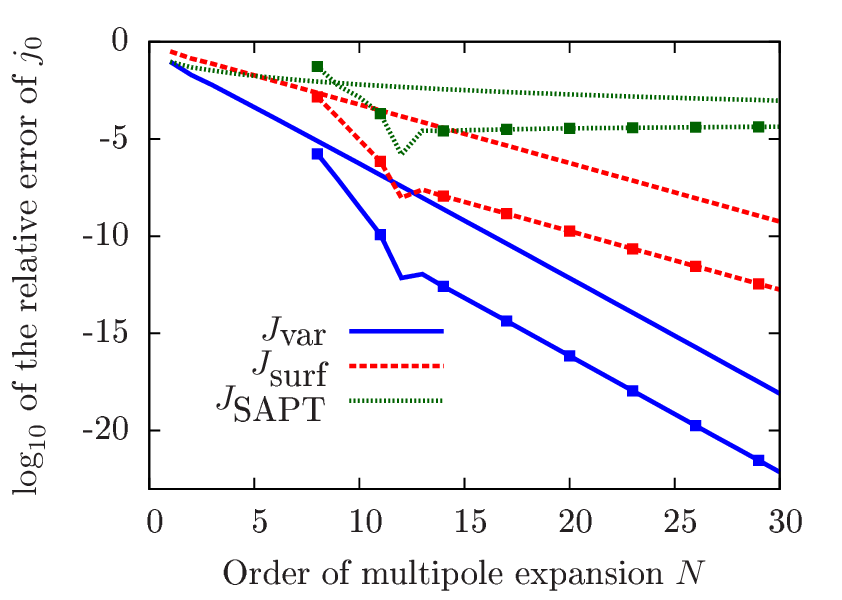}
    \caption{\label{fig:log10_Del_j0_vs_N} (Color online)  
Convergence of the    $j_0$  term of the asymptotic expansion of $J(R)$ obtained 
when the multipole expansion of the  primitive function is used in the variational, 
surface-integral, or SAPT formulas. Lines with squares indicate 
values obtained from eight-term Levin-$u$ extrapolation. }
\end{figure}

\begin{figure}[t!]
  \includegraphics{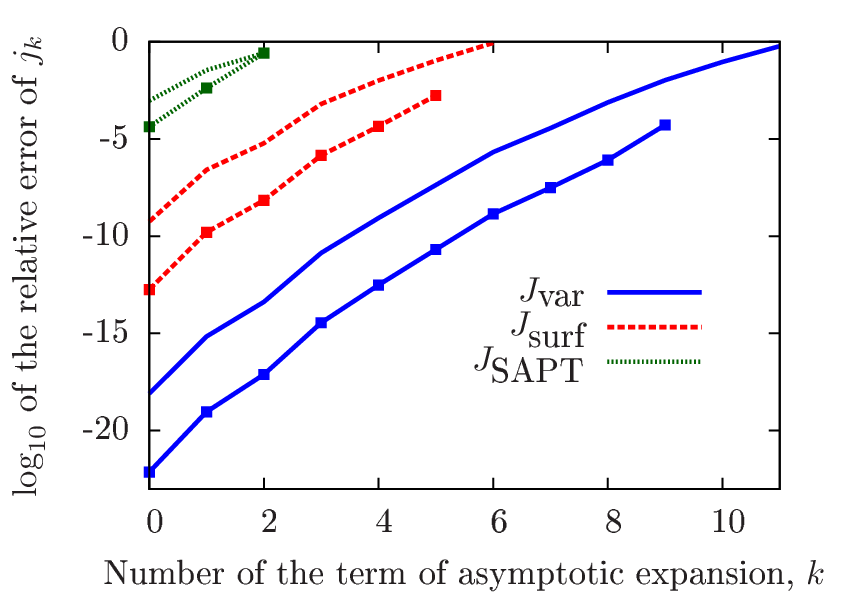}
    \caption{\label{fig:log10_Del_jk_vs_k} (Color online)  
Accuracy of the   $j_k$ coefficients in  the asymptotic expansion of $J(R)$
calculated using different exchange   energy formulas and the multipole expansion 
of the primitive function, Eq. (\ref{eq2}),  summed  through
   $N$=30. Lines with squares indicate 
values obtained from eight-term Levin-$u$ extrapolation.  }
\end{figure}

The regular convergence with respect to $N$ occurs  
not only for $j_0$ but also for higher coefficients, allowing for successful extrapolation. 
 Figure~\ref{fig:log10_Del_jk_vs_k} presents  the decimal logarithms
of relative errors of $j_k$'s calculated with the three exchange 
energy formulas investigated in our work. This graph shows the 
accuracy of the raw   results obtained for $N=30$
and the  very significant gain in accuracy due to  the   Levin   extrapolation.
Note that in Fig.~\ref{fig:log10_Del_jk_vs_k} we did not show 
the results of Levin's $u$-transform for $j_{6}^{\textrm{surf}}$, 
$j_{10}^{\textrm{var}}$, and $j_{11}^{\textrm{var}}$, as in these 
cases extrapolation cannot increase accuracy. 
This is because the $N=30$ terms are not sufficient to establish the regularity of convergence.
Nevertheless, it can be concluded that Levin's  $u$-transform is 
efficient at accelerating the convergence of the series 
investigated in our work, provided adequately large $N$ is used.

\subsection{Exchange splitting from the  multipole expansion of the  first-order  polarization  function}

In practice, the simplest nontrivial approximation to the primitive function $\varphi$ 
is provided by the first-order polarization function $\Phi^{(1)}=\varphi_0 +
\varphi^{(1)}$, defined in Sec.~\ref{subsec:Polarization_expansion_phi}.  In the present subsection we shall find out how accurate 
values  of $j_0$ can be obtained from the multipole expansion of $\varphi^{(1)}$, given by Eq. (\ref{phi1}).   

The value of  $j_0^{\textrm{SAPT}}[\Phi^{(1)}_N]$  has been already given in the literature \cite{Chalasinski:73b}.  
Using our notation the result of Ref.~\onlinecite{Chalasinski:73b}   can be restated  as follows 
\begin{equation}\label{eq:j0_SAPT_psi_N}
 \frac{2}{e} j_0^{\textrm{SAPT}}[\Phi^{(1)}_N] = -\frac{2}{3} -\sum_{n=2}^N \frac{4}{n(n+1)(n+2)(n+3)}.
\end{equation}
The inverse d'Alembert ratio of the  series  on the r.h.s.  of
 Eq.~(\ref{eq:j0_SAPT_psi_N}) is equal to  $1 + 4\,n^{-1} + O(n^{-2}) $ so its convergence radius 
is equal to 1, i.e.,  is the same as for the series of Eq.~(\ref{j0_SAPT}).  It is disappointing that the series of
 Eq.~(\ref{j0_SAPT}),  obtained with the function $\Phi_N$, exhibit somewhat 
poorer  convergence rate that the series  of  Eq.~(\ref{eq:j0_SAPT_psi_N}}), obtained 
with the first-order approximation  $\Phi^{(1)}_N$ to   $\Phi_N$ . 
Taking into account that the sum of the series  in Eq.~(\ref{eq:j0_SAPT_psi_N}) equals    $-1/18$,  we find that   
  $j_0^{\textrm{SAPT}}[\Phi^{(1)}] =  - 13e/36$,
which differs from the exact value by  $-$1.8\%. This result has been obtained 
using different method (without the multipole expansion) in Ref.~\onlinecite{Chalasinski:73a}.

The expression for $j_{\textrm{surf}}[\Phi^{(1)}_N]$ can be deduced from the results 
of Tang et al.\cite{Tang:91} Equation~(7.25) of that work implies that
\begin{equation}\label{eq:j0_surf_psi_N}
 \frac{2}{e} j_0^{\textrm{surf}}[\Phi^{(1)}_N] = - \frac{ 1 }{ 2 } \bigg( 1 +\sum_{n=2}^N \frac{1}{n \,2^n} 
   \bigg )^2 .
\end{equation}
One can easily find that the increment ratio for the sequence  
$j_0^{\textrm{surf}}[\Phi^{(1)}_N] $ equals $2 +  2\, N^{-1} + O(N^{-2})$,
so the corresponding convergence radius is equal 2.    
 The $N\rightarrow \infty$ limit of the r.h.s. of Eq.~(\ref{eq:j0_surf_psi_N})    
is  $-\frac{1}{2} \left ( \ln 2 + \frac{1}{2} \right )^2$, and the corresponding 
approximate value of $j_0$, given by  $j_0^{\textrm{surf}}[\Phi^{(1)}_N] $,
  differs by $-$3.3\% from the exact value $j_0=-1$.
 Thus, the first-order polarization function gives a better approximation to the exchange 
splitting when used in the SAPT formula than in the surface-integral formula.   

The evaluation of $j_0^{\textrm{var}}[\Phi^{(1)}_N] $ is somewhat more complicated.  
Using the large-$R$ asymptotic estimation of the integral 
\begin{equation}\begin{aligned}
 & \frac{1}{\pi}\int e^{-r_a-r_b}  r_a^k r_b^l P_m(\cos\theta_a)P_n(\cos\theta_b)d^3{\bf  r} = \\
& \quad = 2e^{-R} \frac{(k+1)! (l+1)!}{(k+l+3)!}[ R^{k+l+2}  +O(R^{k+l+1})]
\end{aligned}\end{equation}
one can find that 
\begin{equation}\label{eq:j0_var_psi_N}\begin{aligned}
 \frac{2}{e} j_0^{\textrm{var}}[\Phi^{(1)}_N] = -\frac{2}{3} -\sum_{k=2}^N \frac{ 8 }{ k(k+1)(k+2)(k+3) }  \\
 + 2  \sum_{k,l=2}^N \frac{(k-1)! (l-1)! [kl(k+l+4) - 2 ] }{ (k+l+3)! } .
\end{aligned}\end{equation}
The double sum  in Eq.~(\ref{eq:j0_var_psi_N}) and its $N \rightarrow \infty$ limit  can be worked out analytically and  one obtains 
\begin{equation}\label{Chip73}
   \frac{2}{e} \lim_{N \rightarrow \infty} j_0^{\textrm{var}}[\Phi^{(1)}_N] = -\frac{989}{540}+\frac{\pi^2}{9} .
\end{equation}
The resulting approximate value of $j_0$ differs from the exact one by only  0.12\%. 
 It may be noted that the value of   $ j_0^{\textrm{var}}[\Phi^{(1)}]$, equivalent 
to Eq.~(\ref{Chip73}), has been obtained earlier by Chipman and Hirschfelder  \cite{Chipman:73}
without the help of the multipole expansion using the closed-form expression  for $\varphi^{(1)}$. 

\begin{figure}
  \includegraphics{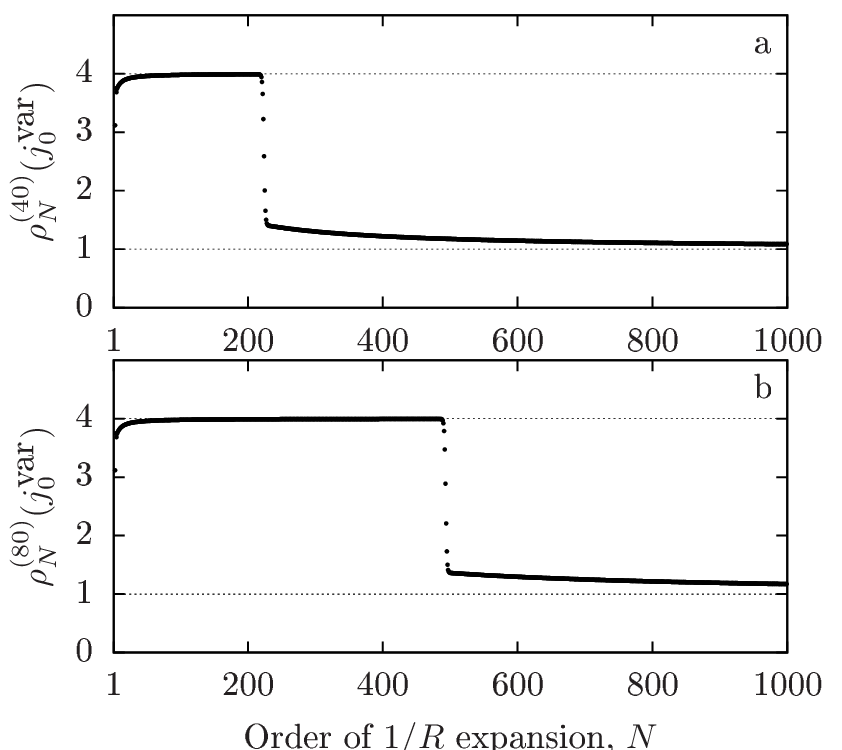}
 \caption{The increment rations  $\rho^{(K)}_N( j_0^{\textrm{var}} )$ for $K=40$ (panel~a) 
 and $K=80$ (panel b) as a function of $N$. } 
 \label{fig:dAlemb_jk_RS} 
 \end{figure}

The $N$ convergence  of the sequence on the r.h.s. of 
Eq.~(\ref{eq:j0_var_psi_N})  turns out to be rather slow. One can show analytically  that the ratio 
of the $N$th  and the  $(N+1)$th increments, defined as in Eq.~(\ref{rhoN}) 
with $\Phi_N$ replaced by $\Phi^{(1)}_N$ and denoted by 
$\rho^{(1)}_N( j_0^{\textrm{var}})$,  behaves at large $N$ as   
 \begin{equation}\label{rho1}
 \rho^{(1)}_N( j_0^{\textrm{var}} ) = 1 + 6\, N^{-1} + O(N^{-2}) .
 \end{equation}
Thus, the sequence $j_0^{\textrm{var}}[\Phi^{(1)}_N]$ converges at large $N$ as a series with the convergence radius equal to 1.  
This seems to be in a disagreement with Eq.~(\ref{eq:rN_jk_var}), which may suggest a faster
convergence.  It turns out, however, that when the multipole expansion of the higher polarization functions 
$\Phi^{(K)}_N$  is used,
the $N$-convergence of  $j_0^{\textrm{var}}[\Phi^{(K)}_N]$  corresponds also to the convergence radius 1. 
We have found by fitting the following large-$N$ behavior of  $\rho^{(K)}_N( j_0^{\textrm{var}} ) $  
\begin{equation}\label{rhoK}\begin{aligned}
 \rho^{(K)}_N & ( j_0^{\textrm{var}} ) = 1 +  ( 2K + 4 ) \, N^{-1} \\
 & - (K-1) N^{-1} ( \ln N )^{-1} +  O \big ( N^{-1} (\ln N)^{-2} \big ) .
\end{aligned}\end{equation}
 The prefactor $2K+4$ in the subleading term in Eq~(\ref{rhoK}) 
shows that although the convergence radius for each $K$ is equal to 1, the rate of convergence improves with increasing $K$. 
This rate becomes  geometric only in infinite order in $V$, i.e., at  $K=\infty$, when Eq.~(\ref{eq:rN_jk_var})
holds. However, we have noted  in Sec.~\ref{subsec:Polarization_expansion_phi} 
that  $\Phi^{(K)}_N=\Phi_N$  when $N\le  2K$. One can expect, then,  that    
\begin{equation}\label{eq:rN_j0_var_Phi_p}
\rho^{(K)}_N( j_0^{\textrm{var}} )  = \rho_N( j_0^{\textrm{var}} ) = 4 - 2 \, N^{-1} + O(N^{-2})  ,
\end{equation}
in the range of  $N$ values smaller or equal  to  2$K$. Thus,    $\rho^{(K)}_N( j_0^{\textrm{var}} )$ is given by
Eq.~(\ref{rhoK}) at $N \!\gg\! 2K$ and by Eq~(\ref{rhoK}) at $N\le  2K$. This is indeed the case as shown in Fig.~\ref{fig:dAlemb_jk_RS}, where 
the increment rations $\rho^{(K)}_N( j_0^{\textrm{var}} )$ are presented for $K$=40 and $K$=80.   
The switch from the fast geometric convergence at low $N$ to the slow harmonic convergence 
at high $N$ is well seen. 
One can show\cite{Gniewek:15} that this switch occurs at $N\approx N_c$, where $N_c$ is the solution of a transcendental equation 
\begin{equation}
 \frac{ \Gamma( N_c+1 ) \Gamma( N_c+2K+6 ) }{ \Gamma( 2N_c+3 ) } = 2 e \frac{ (2K+2)! }{ 2^K K! } .
\end{equation}
This equation correctly describes the behavior of $\rho_N^{(K)}$, for instance for $K=40$ and $K=80$ 
it gives $N_c = 232.3$ and $K=80$ $N_c = 507.9$, respectively, in a good agreement with the values deduced from Fig.~\ref{fig:dAlemb_jk_RS}.
The conclusion of the considerations in this subsection is that 
the inclusion of high-order effects in $V$ is necessary to obtain a fast converging approximation of the exchange splitting energy at large $R$.

\subsection{Exchange splitting from a variational approximation 
to the primitive function}

The calculations of high-order multipole corrections 
may  not be practical  for many-electron systems, nevertheless we believe that
the methods described above can be adapted 
for larger  diatomics. This can be achieved via the use of
Rayleigh-Ritz variational calculations of the  primitive function with appropriately 
 localized basis
set, i.e.,  with trial   functions  satisfying the  localization 
condition of Eq.~(\ref{local}). 
In order to test the effectiveness of this approach  we applied the  basis set  
employed  in our previous work on the same system \cite{Gniewek:14}, but  
restricted to orbitals centered exclusively at the nucleus $a$. 
This restriction makes the basis set orthogonal for any $R$.
The parameter $\Omega$   employed  in Ref. \onlinecite{Gniewek:14} to control the size of the basis set 
(defined as the  maximal sum of the order of the included Laguerre and Legendre polynomials)
is closely connected to the extent of the multipole expansion of $\varphi$ that can be recovered by the variational calculation. 
For  example $\varphi_{N}$ can be recovered  with the $\Omega = N$, 
but not with the  $\Omega = N-1$ basis set. 
We performed variational calculations of $\varphi$ for 46 internuclear distances $R$=60, 62,\ldots,150, followed  
by the application of the exchange energy formulas $J_{\textrm{SAPT}}[\varphi]$, $J_{\textrm{surf}}[\varphi]$
and $J_{\textrm{var}}[\varphi]$   and least-squares fitting 
of the obtained values of $J(R)$ to extract the constants $j_0$, $j_1$ and $j_2$.  
The results of these calculations for $\Omega=10$ are given in Table~\ref{tab:mult_vs_VPLB} together 
with the values obtained with  the function $\Phi_{10}$, which are   given for 
comparison. 
It is seen that the agreement of the $j_0$ values
calculated with the multipole expansion and with the variational approximation to $\varphi$ 
is excellent. In the case of  the higher constants $j_1$ and $j_2$ the agreement is not so
good but reasonable, the multipole expansion giving consistently better results. 
As expected the volume-integral formula of Eq.~(\ref{eq:J_var}) performs best not only in the case of 
the multipole expansion of the primitive function, but  also with the variational 
approximation to this function.

\begin{table}[t]
\caption{\label{tab:mult_vs_VPLB} Values of $j_0$, $j_1$ and $j_2$
 evaluated using the surface- and volume-integral 
formulas and  the  primitive function approximated using either 
the multipole expansion through the $10$th order ($\Phi_{10}$),  or  the  variational  
treatment with localized, $\Omega=10$ basis  ($\Phi^{V\!L\!B}_{10}$).   
 }
\begin{ruledtabular}
\begin{tabular}{ l d{2.12} d{2.7} d{2.7}  } 
method &  \multicolumn{1}{c}{$j_0$} &   \multicolumn{1}{c}{$j_1$} &   \multicolumn{1}{c}{$j_2$} \\[0.5ex]
\hline\\ [-1.5ex]
$J_{\textrm{SAPT}}[\Phi_{10}]$            &  -0.9935\ 8974   &  -0.444\ 639  &  2.2909 \\[0.5ex]
$J_{\textrm{SAPT}}[\Phi^{V\!L\!B}_{10}]$  &  -0.9935\ 8952   &  -0.490\ 467  &  1.8381 \\[0.5ex]
$J_{\textrm{surf}}[\Phi_{10}]$            &  -0.9994\ 0777   &  -0.515\ 396  &  3.3341 \\[0.5ex]
$J_{\textrm{surf}}[\Phi^{V\!L\!B}_{10}]$  &  -0.9994\ 0757   &  -0.538\ 712  &  3.8320 \\[0.5ex]
$J_{\textrm{var}}[\Phi_{10}]$             &  -0.9999\ 9946   &  -0.500\ 022  &  3.1260 \\[0.5ex]
$J_{\textrm{var}}[\Phi^{V\!L\!B}_{10}]$   &  -0.9999\ 9965   &  -0.499\ 994  &  3.1185 \\[0.5ex]
exact                                     &  -1.0            &  -0.5         & 3.125
\end{tabular}
\end{ruledtabular}
\end{table}

\section{\label{sec:conclusions} Summary and Conclusions}

Tang et al.\cite{Tang:91} showed that leading constant $j_0$ in the asymptotic 
expansion of the exchange splitting 
$J(R)  =  2 e^{-R-1} R ( j_0 + j_1  R^{-1} + j_2  R^{-2} +\cdots ) $  for H$^+_2$ can be obtained when the surface-integral formula 
$J_{\textrm{surf}}[\varphi]$ of Eq. (\ref{eq:J_surf}) is evaluated with 
 the multipole  expanded polarization series for the wave function. 
These authors inferred that the polarization series 
converges to the primitive function  $\varphi$, rather than to the fully symmetric ground 
state function $\psi_g$, but his conclusion was later shown 
to be invalid, as \'Cwiok \emph{et. al} \cite{Cwiok:92:pol}  gave compelling evidence of the convergence to $\psi_g$.

Our work significantly extends the results of Tang et al.\cite{Tang:91}  
as we have calculated not only the leading but also higher terms in the asymptotic expansion of $J(R)$. 
We applied the surface-integral formula as well, but also two other 
exchange energy expressions that have not been previously considered in the present  context: 
the  volume-integral formula  $ J_{\textrm{SAPT}}[\varphi]$, Eq.~(\ref{SAPT}), 
and the powerful volume-integral formula $ J_{\textrm{var}}[\varphi] $ 
based on the Rayleigh-Ritz variational principle, Eq. (\ref{eq:J_var}). 
We have also calculated the convergence
radii for the expansions resulting when of the constants  $j_0$,$\ldots$,$j_3$ 
are calculated using the considered exchange energy expressions and  
the multipole expansion for the primitive function $\varphi$.  

We found that the variational and surface-integral formulas lead to convergent
expansions for the $j_k$ constants. 
The best convergence and the largest convergence radius, equal to 4, was found  
for the expansions obtained using the variational volume-integral formula.
The convergence radius corresponding to the application  of the surface-integral
formula, $J_{\textrm{surf}}[\varphi]$, was  found to be equal to 2. 
In the case of the SAPT formula, 
we found that the convergence for the constants $j_0$, $j_1$, and $j_2$ 
is slow with the convergence radius equal to 1. Moreover
the expansion for $j_2$ converges to an incorrect value. The expansions for the 
higher constants $j_k$, $k >2$, generated by the SAPT formula turned out to be  divergent.     

When the multipole expansion for the first- or any finite-order polarization 
function is used to represent  the primitive function $\varphi$, the resulting 
expansion for $j_0$ has convergence radius equal to one. Nevertheless the rate 
of convergence improves significantly with increasing order of the polarization 
expansion. This shows the importance of a high-order treatment in the interaction 
potential $V$ to obtain a good approximation for the $j_0$ constant.  
Our results do not contradict the convergence of the polarization expansion 
to the symmetric ground-state function $\psi_g$ (for which all three energy 
formulas are singular) but rely on the fact that the polarization
expansion converges asymptotically to the primitive function $\varphi$.  
 
Since the calculation of high-order multipole corrections for many-electron
systems is not practical, we have presented an alternative method of 
obtaining the primitive function: variational calculation with appropriately   
localized basis set.  We have shown that this variationally obtained primitive function
provides excellent values of $j_0$ and reasonably good approximations 
to higher-order $j_k$ constants. Also in this case the variational volume formula 
provides the most accurate results. 
 
Our study shows that the conventional SAPT formula  exhibits some deficiencies  
for the calculation of the exchange energy at large interatomic distances  of $R$.
The variational volume formula leads to much faster convergence and significantly 
more accurate results when applied both with the multipole expansion of the wave function 
and with a suitable variational approximation to this function.
Moreover this formula provides a much better basis set convergence  of the 
results than the surface integral formula. One can therefore conclude that  
the variational volume-integral formula  provides an attractive alternative 
for the determination of the exchange splitting and the exchange contribution 
of the interaction potential in general.

\begin{acknowledgments}
The authors acknowledge helpful discussions with 
Robert Moszy{\'n}ski.
This work was supported by the 
National Science Centre, Poland, 
project number 2014/13/N/ST4/03833.
\end{acknowledgments}

\bibliography{ref}

\end{document}